# Parameterized Relativistic Dynamical Formalism for Transitions between Three Flavor States

John R. Fanchi
j.r.fanchi@tcu.edu
Department of Engineering and TCU Energy Institute
Texas Christian University
Fort Worth, Texas, 76129 USA

April 4, 2016

## Abstract

Parameterized relativistic dynamics (PRD) is a manifestly covariant quantum theory with invariant evolution parameter. The theory has been applied to neutrino flavor oscillations between two mass states. It is generalized here to transitions between three mass states and applied to electron neutrino oscillations.

## 1. Introduction

Experiments with solar neutrinos, atmospheric neutrinos, reactor neutrinos, and accelerator neutrinos have demonstrated that flavor mixing can occur between two or three neutrino flavors composed of up to three neutrino mass states [Olive, et al, 2014; Gonzalez-Garcia, 2014]. Mass state transitions are a key feature of parametrized relativistic dynamics (PRD). PRD is a manifestly covariant quantum theory invariant evolution parameter. The invariant evolution parameter concept was introduced by Fock [1937] and Stueckelberg [1941, 1942], and later used by Feynman [1948, 1950, 1951] in his path integral formulation of quantum theory in the late 1940's and early 1950's. An overview of PRD is presented by Horwitz [2015], Fanchi [1993, 2011], and Pavsic [2001]. The PRD formalism for two-state flavor mixing [Fanchi, 1998] is extended here to three-state flavor mixing and applied to the neutrino transition $v_e \to v_\mu$ in vacuum.

## 2. Mass Basis and Flavor Basis

We are interested in developing a formalism within the context of PRD that can describe transitions between three neutrino flavor states $\{|v_\alpha\rangle; \alpha = e, \mu, \tau\}$ given the assumption that neutrinos are composed of up to three mass states $\{|v_j\rangle; j = 1, 2, 3\}$. The mass and flavor states

can be written as 3-component column vectors:

$$|v_j\rangle = \begin{bmatrix} |v_1\rangle \\ |v_2\rangle \\ |v_3\rangle \end{bmatrix} \quad (2.1)$$

and

$$|v_\alpha\rangle = \begin{bmatrix} |v_e\rangle \\ |v_\mu\rangle \\ |v_\tau\rangle \end{bmatrix} \quad (2.2)$$

The mass basis $\{|v_j\rangle; j = 1, 2, 3\}$ is related to the flavor basis $\{|v_\alpha\rangle; \alpha = e, \mu, \tau\}$ by a unitary transformation:

$$\begin{bmatrix} |v_e\rangle \\ |v_\mu\rangle \\ |v_\tau\rangle \end{bmatrix} = U \begin{bmatrix} |v_1\rangle \\ |v_2\rangle \\ |v_3\rangle \end{bmatrix} \quad (2.3)$$

where U is the unitary matrix

$$U = \begin{bmatrix} u_{e1} & u_{e2} & u_{e3} \\ u_{\mu 1} & u_{\mu 2} & u_{\mu 3} \\ u_{\tau 1} & u_{\tau 2} & u_{\tau 3} \end{bmatrix} \quad (2.4)$$

satisfying

$$U^{-1} = (U^*)^T \quad (2.5)$$

The elements of the unitary matrix are

$$u_{\alpha j}^{-1} = u_{j\alpha}^*; \ j = 1, 2, 3 \text{ and } \alpha = e, \mu, \tau \quad (2.6)$$

The expanded form of the unitary transformation is

$$\begin{aligned} |v_e\rangle &= u_{e1}|v_1\rangle + u_{e2}|v_2\rangle + u_{e3}|v_3\rangle \\ |v_\mu\rangle &= u_{\mu 1}|v_1\rangle + u_{\mu 2}|v_2\rangle + u_{\mu 3}|v_3\rangle \\ |v_\tau\rangle &= u_{\tau 1}|v_1\rangle + u_{\tau 2}|v_2\rangle + u_{\tau 3}|v_3\rangle \end{aligned} \quad (2.7)$$

A mass basis state satisfies the temporal evolution equation

$$\begin{aligned} T|v_j\rangle &= i\hbar \frac{\partial}{\partial s}|v_j\rangle \\ &= T_j |v_j\rangle \\ &= \hbar^2 \frac{k_j^\mu k_{j\mu}}{2m_j}|v_j\rangle \end{aligned} \quad (2.8)$$

where $T_j$ is the eigenvalue of the temporal operator $T$, $s$ is the scalar evolution parameter, $k_j^\mu$ is the energy-momentum of state $j$, and $m_j$ is the mass of state $j$. Equation (2.8) has the formal solution

$$\begin{bmatrix} |v_1\rangle \\ |v_2\rangle \\ |v_3\rangle \end{bmatrix} = \begin{bmatrix} e^{-i\frac{T_1 s}{\hbar}} & 0 & 0 \\ 0 & e^{-i\frac{T_2 s}{\hbar}} & 0 \\ 0 & 0 & e^{-i\frac{T_3 s}{\hbar}} \end{bmatrix} \begin{bmatrix} |v_1(0)\rangle \\ |v_2(0)\rangle \\ |v_3(0)\rangle \end{bmatrix} \qquad (2.9)$$

where $v_j(0)$ is mass state $j$ at $s = 0$.

## 3. Transitions between Flavor States: Electron Neutrino Disappearance

The formalism for three mass states presented in the previous section is illustrated by applying the formalism to the disappearance of electron neutrinos. We begin with a pure beam of electron neutrinos in flavor state $|v_e\rangle$. The probabilities of forming $v_\mu$ and $v_\tau$ are

$$P(v_e \to v_\mu) = |\langle v_\mu | v_e(s) \rangle|^2 \qquad (3.1)$$

and

$$P(v_e \to v_\tau) = |\langle v_\tau | v_e(s) \rangle|^2 \qquad (3.2)$$

respectively. The matrix element for $\beta = \mu$ or $\tau$ is

$$\langle v_\beta | v_e(s) \rangle = [u_{\beta 1}^* v_1^* + u_{\beta 2}^* v_2^* + u_{\beta 3}^* v_3^*] \times [u_{e1} v_1(s) + u_{e2} v_2(s) + u_{e3} v_3(s)] \qquad (3.3)$$

or, in expanded form,

$$\begin{aligned}\langle v_\beta | v_e(s) \rangle &= u_{\beta 1}^* [u_{e1} v_1^* v_1(s) + u_{e2} v_1^* v_2(s) + u_{e3} v_1^* v_3(s)] \\ &+ u_{\beta 2}^* [u_{e1} v_2^* v_1(s) + u_{e2} v_2^* v_2(s) + u_{e3} v_2^* v_3(s)] \\ &+ u_{\beta 3}^* [u_{e1} v_3^* v_1(s) + u_{e2} v_3^* v_2(s) + u_{e3} v_3^* v_3(s)]\end{aligned} \qquad (3.4)$$

Equation (3.4) is simplified by applying the orthonormality condition $\langle v_i | v_j \rangle = \delta_{ij}$ to obtain

$$\begin{aligned}\langle v_\beta | v_e(s) \rangle &= u_{\beta 1}^* u_{e1} v_1^* v_1(s) \\ &+ u_{\beta 2}^* u_{e2} v_2^* v_2(s) \\ &+ u_{\beta 3}^* u_{e3} v_3^* v_3(s)\end{aligned} \qquad (3.5)$$

The temporal dependence is obtained by expressing Eq. (3.5) in terms of the mass state at $s = 0$:

$$\langle v_\beta | v_e(s) \rangle = u^*_{\beta 1} u_{e1} \exp\left(-i\frac{T_1 s}{\hbar}\right)$$
$$+ u^*_{\beta 2} u_{e2} \exp\left(-i\frac{T_2 s}{\hbar}\right) \quad (3.6)$$
$$+ u^*_{\beta 3} u_{e3} \exp\left(-i\frac{T_3 s}{\hbar}\right)$$

## 4. Application to the $v_e \to v_\mu$ Transition

We apply the 3 flavor state-formalism to the transition $v_e \to v_\mu$ in vacuum between two flavor states. In this application we assume the transition $v_e \to v_\tau$ is negligible and that only 2 flavor states are involved. A 3-state unitary matrix that effectively simplifies the problem so that we only need to consider flavor states 1 and 2 is

$$U(3) = \begin{bmatrix} U_{12}(2) & 0 \\ 0 & 1 \end{bmatrix}$$
$$= \begin{bmatrix} \cos\theta_{12} & \sin\theta_{12} & 0 \\ -\sin\theta_{12} & \cos\theta_{12} & 0 \\ 0 & 0 & 1 \end{bmatrix} \quad (4.1)$$

where $\theta_{12}$ refers to the mixing angle between mass states 1 and 2 in vacuum.

The transition probability amplitude is

$$\langle v_\mu | v_e(s) \rangle = u^*_{\mu 1} u_{e1} \exp\left(-i\frac{T_1 s}{\hbar}\right)$$
$$+ u^*_{\mu 2} u_{e2} \exp\left(-i\frac{T_2 s}{\hbar}\right) \quad (4.2)$$

where $\{T_j, j = 1,2\}$ are the eigenvalues of the temporal evolution operator, and

$$u^*_{\mu 1} = -\sin\theta_{12}$$
$$u_{e1} = \cos\theta_{12}$$
$$u^*_{\mu 2} = \cos\theta_{12} \quad (4.3)$$
$$u_{e2} = \sin\theta_{12}$$

Substituting Eq. (4.3) into Eq. (4.2) gives

$$\langle v_\mu | v_e(s) \rangle = -\sin\theta_{12} \cos\theta_{12} \exp\left(-i\frac{T_1 s}{\hbar}\right)$$
$$+ \cos\theta_{12} \sin\theta_{12} \exp\left(-i\frac{T_2 s}{\hbar}\right) \quad (4.4)$$
$$= \sin\theta_{12} \cos\theta_{12} \left[\exp\left(-i\frac{T_2 s}{\hbar}\right) - \exp\left(-i\frac{T_1 s}{\hbar}\right)\right]$$

and

$$\langle v_\tau | v_e(s) \rangle = 0 \quad (4.5)$$

The transition probability is

$$P(v_e \to v_\mu) = |\langle v_\mu | v_e(s) \rangle|^2 \quad (4.6)$$

## 5. Application to Neutrino Oscillation Experiments

The evolution equation in PRD for a state may be written in terms of the evolution parameter $s$ as

$$i\hbar \frac{\partial}{\partial s} |v_j\rangle = K_j |v_j\rangle \quad (5.1)$$

where $K_j$ is the eigenvalue of the mass operator for mass state $j$. The evolution parameter dependent solution of Eq. (5.1) in the mass basis for two mass states is

$$\begin{bmatrix} |v_1(s)\rangle \\ |v_2(s)\rangle \end{bmatrix} = \begin{bmatrix} e^{-iK_1 s/\hbar} & 0 \\ 0 & e^{-iK_2 s/\hbar} \end{bmatrix} \begin{bmatrix} |v_1(0)\rangle \\ |v_2(0)\rangle \end{bmatrix} \quad (5.2)$$

where

$$K_j = \hbar^2 k_j^\mu k_{j\mu}/2m_j = \hbar^2 \left[(\omega_j/c)^2 - k_j \cdot k_j\right]/2m_j \quad (5.3)$$

In PRD, the components of the energy-momentum four-vector $k_j^\mu$ are observables and the mass $m_j$ is a function of statistical values of $k_j^\mu$.

In the flavor oscillation process $v_e \to v_\mu$, we begin with a pure beam of electron neutrino $v_e$ particles and calculate the probability for formation of muon neutrino $v_\mu$ particles. The PRD result for the probability of forming the final state $v_\mu$ from initial state $v_e$ is

$$P_{PRD}(v_e \to v_\mu) = \sin^2 2\theta \sin^2\left\{\frac{(m_2 - m_1)c^2}{4\hbar} s\right\} \quad (5.4)$$
$$\equiv \sin^2 2\theta \sin^2 \alpha_{PRD}$$

where $s$ is temporal duration measured by an evolution parameter clock [Fanchi, 2011]. Dynamical factors are collected in the term $\alpha_{PRD}$.

The value of the invariant evolution parameter $s$ is determined in PRD by introducing an s-clock. In our application, flavor oscillations are described by quantifying the behavior of two particles. One particle propagates without interaction or oscillation from the source to the detector and serves as a "clock" for the scalar evolution parameter s. The other particle is the oscillating particle. In this application, the source and detector are separated by a distance $L$.

The most probable trajectory of the non-interacting s-clock particle is

$$s^2 = (\delta t)^2 - \frac{(\delta x)^2}{c^2} = (\delta t)^2 [1 - \beta^2], \beta = \frac{v}{c}, v = \frac{\delta x}{\delta t} \tag{5.5}$$

The distance $\delta x$ traveled by the s-clock particle in the interval $\delta t$ is $L$, so we obtain

$$s = \frac{L}{c} \frac{[1-\beta^2]^{1/2}}{\beta}, \delta x = L \tag{5.6}$$

Substituting Eq. (5.6) into Eq. (5.4) gives

$$P_{PRD}(v_e \to v_\mu) = \sin^2 2\theta \sin^2 \alpha_{PRD},$$

$$\alpha_{PRD} = \frac{(m_2 - m_1)c^2}{4\hbar} \frac{L}{c} \frac{[1-\beta^2]^{1/2}}{\beta} \tag{5.7}$$

The result for the conventional theory denoted by subscript $Std$ is

$$P_{Std}(v_e \to v_\mu) = \sin^2 2\theta \sin^2 \alpha_{Std}$$

$$\alpha_{Std} = \frac{(m_2^2 - m_1^2)c^4}{4\hbar} \frac{L}{cE_v} \tag{5.8}$$

where $E_v$ is the energy of the ultrarelativistic incident neutrino

$$E_v = \frac{m_v c^2}{[1-\beta^2]^{1/2}} \tag{5.9}$$

We combine Eqs. (5.7) and (5.9) and rearrange to simplify comparison with Eq. (5.8):

$$P_{PRD}(v_e \to v_\mu) = \sin^2 2\theta \sin^2 \alpha_{PRD},$$

$$\alpha_{PRD} = \frac{(m_2 - m_1)c^2}{4\hbar} \frac{L}{c} \frac{m_v c^2}{E_v \beta} = \frac{m_v (m_2 - m_1)c^4}{4\hbar E_v} \frac{L}{c} \frac{1}{\beta} \tag{5.10}$$

The ratio of the dynamical factors $\alpha_{PRD}, \alpha_{Std}$ is

$$\frac{\alpha_{Std}}{\alpha_{PRD}} = \frac{m_2^2 - m_1^2}{m_v (m_2 - m_1)} \beta = \frac{m_1 + m_2}{m_v} \beta \tag{5.11}$$

and the ratio of probabilities in Eqs. (5.8) and (5.10) is

$$\frac{P_{Std}}{P_{PRD}} = \frac{\sin^2 \alpha_{Std}}{\sin^2 \alpha_{PRD}} \tag{5.12}$$

Comparing P$_{PRD}$, P$_{Std}$ and the dynamical factors $\alpha_{PRD}, \alpha_{Std}$ shows that the PRD model and the conventional theory have the same dependence on the flavor mixing angle θ, but their dependence on dynamical factors differs significantly. If the mass difference between neutrino mass and flavor states is very small and the neutrinos are ultrarelativistic, then $(m_1 + m_2)/m_v \approx 2$ since $\beta \approx 1$ and $m_1 \approx m_2 \approx m_\nu$. The ratio of dynamical factors is $\alpha_{Std}/\alpha_{PRD} \approx 2$ in this case.

The survival probability of the electron neutrino is

$$P_{PRD}(\nu_e \to \nu_e) = 1 - P_{PRD}(\nu_e \to \nu_\mu) = 1 - \sin^2 2\theta \sin^2 \alpha_{PRD} \qquad (5.13)$$

in the PRD model, and

$$P_{Std}(\nu_e \to \nu_e) = 1 - P_{Std}(\nu_e \to \nu_\mu) = 1 - \sin^2 2\theta \sin^2 \alpha_{Std} \qquad (5.14)$$

in the conventional Std model. The survival probabilities agree with Rusov and Vlasenko [2012]. They used the relationship $\alpha_{Std}/\alpha_{PRD} \approx 2$ and available data in a postulated mass matrix for 3 mass states to estimate neutrino masses.

Survival probabilities for the conventional (Std) model and the PRD model are compared in Figure 1. The angle $\alpha_{Std}$ is calculated using $L = 180\,km$, $\Delta m^2 = 7.0 \times 10^{-5} eV^2$ and $\sin^2 2\theta = 0.84$ as a function of neutrino energy that varies from 0.1 MeV to 15 MeV [Olive, et al., 2014, Fig. 14.1, pg. 63; and Goswami, et al., 2005]. The angle $\alpha_{PRD}$ is calculated using $\alpha_{Std}/\alpha_{PRD} \approx 2$ which is based on the assumption that the neutrinos are ultrarelativistic in vacuum and $(m_1 + m_2)/m_\nu \approx 2$.

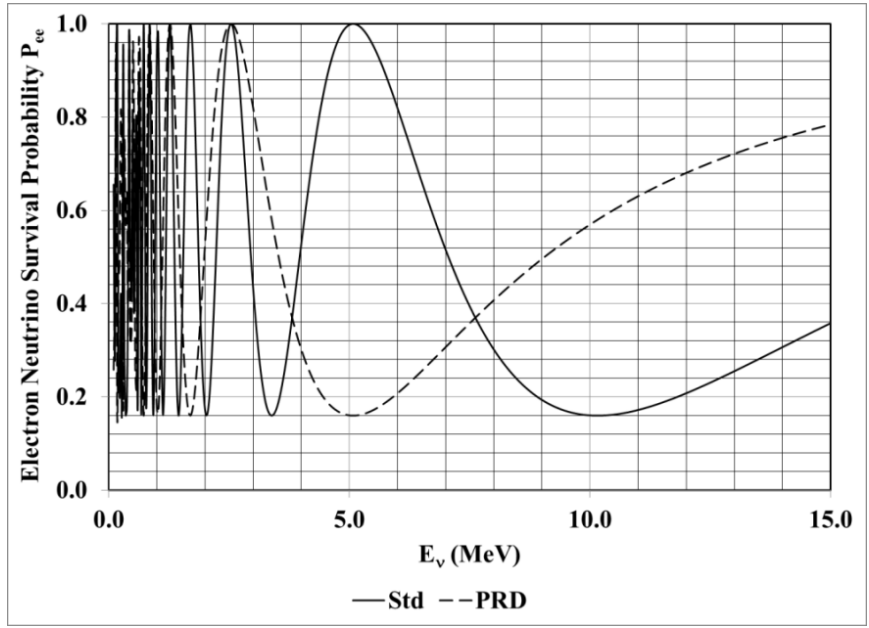

**Figure 1. Comparison of Conventional (Std) and PRD Survival Probabilities as a Function of Neutrino Energy**

Figure 2 shows survival probability of the electron neutrino as a function of $L$ at a neutrino energy of 10 MeV. The angle $\alpha_{Std}$ is calculated using $\Delta m^2 = 7.0 \times 10^{-5} eV^2$ and $\sin^2 2\theta = 0.84$.

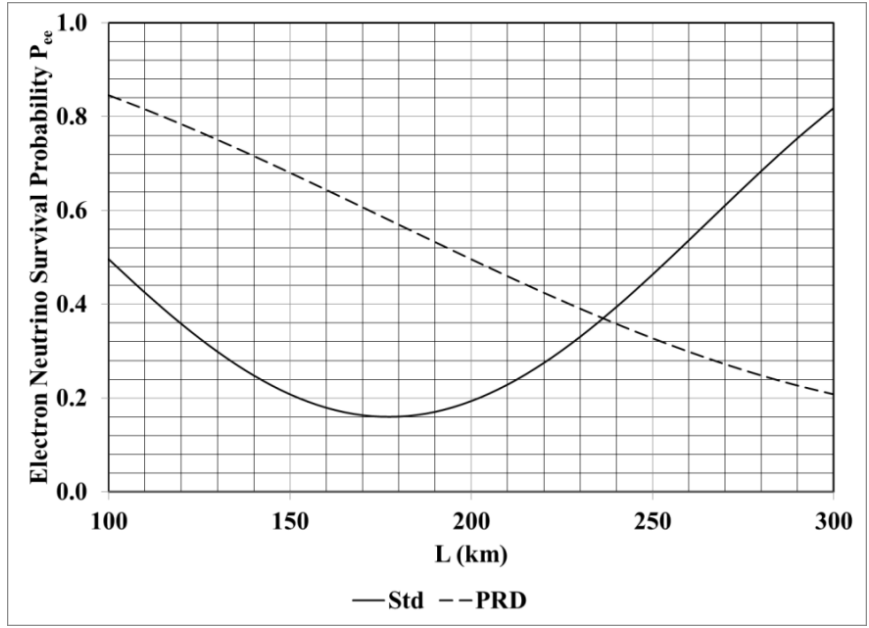

**Figure 2. Comparison of Conventional (Std) and PRD Survival Probabilities as a Function of the Distance $L$ between Source and Detector**

It is clear from the figures that there are significant differences between conventional (Std) and PRD theoretical results. The display of experimental results should provide theory-independent information that can be used to determine the probability of disappearance $P(v_e \to v_\mu)$ or survival $P(v_e \to v_e)$ of electron neutrinos. Further work will be needed to examine experimental results within the context of PRD. The result may be a set of neutrino masses that is consistent with the experimental results but differs from the conventional analysis.

## 6. Conclusions

A formalism for studying mass state transitions between neutrino flavor states was presented in the context of parameterized relativistic dynamics (PRD) and applied to the survival of electron neutrinos. The analysis shows that significant differences exist between theoretical results of the conventional model and the PRD model.